# pH-dependent interfacial rheology of polymer membranes assembled at liquid-liquid interfaces using hydrogen bonds

Julien Dupré de Baubigny[1], Corentin Trégouët[1,2,5], Elena Govorun[2,3], Mathilde Reyssat[2], Patrick Perrin[1], Nadège Pantoustier[1], Thomas Salez[4,*], Cécile Monteux[1,*]

[1] Laboratoire Sciences et Ingénierie de la Matière Molle, CNRS UMR 7615, ESPCI Paris, PSL Research University, Sorbonne Université, 10 rue Vauquelin, Paris, France.

[2] Laboratoire Gulliver, CNRS UMR 7083, ESPCI Paris, PSL Research University, Sorbonne Université, 10 rue Vauquelin, Paris, France.

[3] Laboratoire de Physique Théorique et Modèles Statistiques, CNRS UMR 8626, University Paris-Saclay, Orsay, France.

[4] Univ. Bordeaux, CNRS, LOMA, UMR 5798, 33405 Talence, France.

[5] Chimie Biologie Innovation, CNRS, ESPCI Paris, PSL Research University.

**Abstract**

Self-assembly of polymers at liquid interfaces using non-covalent interactions has emerged as a promising technique to reversibly produce self-healing membranes. Besides the assembly process, it is also crucial to control the mechanical properties of these membranes. Here, we measure the interfacial rheological properties of PMAA-PPO (polymethacrylic acid - polypropylene oxide) polymer membranes assembled using hydrogen bonds at the interface between water and a polar oil, Mygliol. Varying the pH enables us to modify the degree of ionization of the PMAA chains, and hence their ability to establish hydrogen interactions with PPO. Frequency sweeps of the interfacial layers show a crossover between a viscous regime at low frequencies and an elastic regime at high frequencies. The crossover elastic modulus, measured one hour after the two phases were put into contact, decreases by a half over the pH range investigated, which can be accounted for by a decrease of the layer thickness as pH increases. Furthermore, we find that the crossover frequency varies exponentially with the degree of ionization of PMAA. To account for these observations, we propose a simple picture where the short PPO chains behave as non-covalent cross-linkers that bridge several PMAA chains. The dissociation rate and hence the crossover frequency are controlled by the number of PO units per PPO chain involved in the hydrogen bonds.

## Introduction

Assembling polymer chains at liquid interfaces has emerged as an easy method to obtain self-healing and reconfigurable membranes for applications in encapsulation or biomedicine [1]. To obtain interfacial membranes, one can use non-covalent interactions between two polymers located in two immiscible liquid phases [2-9], between a polymer and a surfactant [10-14], or between a polymer and nanoparticles [15-17]. One can also trigger the assembly of a membrane at the interface between two aqueous solutions such as in alginate – calcium systems where divalent calcium ions bridge biopolymer molecules [18-19]. Interfacial polymer membranes have also been obtained at the interface of two immiscible polymer solutions in a case where the continuous phase is water, such as in phase separating PEG-Dextran solutions known as water-water interfaces [3, 4, 6]. In the latter case, the interactions used to obtain an interfacial membrane were mainly electrostatic [2-7, 9-19] but hydrogen bonds have also been employed in the same context [8, 20, 21]. To understand the growth of such interfacially-assembled membranes, the evolution of the membrane thickness, h, as a function of time, t, was measured by several groups. They reported either a thickness that grows as $h \sim t^{1/2}$ [8, 11, 20] or an exponential growth with time [14]. In the case where a $h \sim t^{1/2}$ growth was observed, the assembly process is controlled by the diffusion of one of the species through the membrane. In previous works, we studied the growth of interfacial membranes between PMAA and PPO (polymethacrylic acid and polypropylene oxide) through hydrogen bonds. Indeed, the metacrylic acid units of the PMAA chains establish hydrogen bonds with the propylene oxide units of the PPO chains. We suggested that the growth of these membranes is controlled by the diffusion of PPO chains through the growing membrane, and that the diffusion coefficient of the PPO chains decreases with the PPO concentration as the PPO chains have to stretch in order to diffuse through the crowded membrane [20]. More recently, we demonstrated that pH controls membrane growth as it determines the PMAA ionization degree and the concentration of counterions, and hence the number of hydrogen bonds, the gel composition, and the adsorption rate of charged PMAA chains at the membrane surface [21]. Looking at the assembly process at early times for a mixture of chitosan and an oppositely-charged surfactant, using light scattering, de Loubens et al. showed the formation of aggregates whose number grows over time and who finally form a percolated network at the interface [22]. The formation of the network was confirmed by the same authors using interfacial rheological measurements. In fact, controlling the mechanical properties of such membranes is crucial as these properties control their resistance to shear or elongation when transported and deformed in a flow [23, 24].

In this article, we investigate the interfacial rheology of PMAA-PPO layers as a function of pH, and hence the degree of ionization of PMAA chains. We find that the elastic interfacial modulus of the interfacial layer can be decreased by a factor two by increasing the pH from 3 to 5 and that this effect can be accounted for by a simple variation of thickness of the layers. Furthermore, the crossover frequency of the G' and G'' frequency curves increases exponentially with the degree of ionization. To account for these results, we propose a simple picture where the PMAA chains are bridged by PPO chains and where the crossover frequency is controlled by the probability of PPO chains to detach all the hydrogen bonds that they establish with PMAA chains.

**Materials and methods**

A schematic of the system is shown in Figure 1. Aqueous PMAA solutions are prepared by dissolving 1 wt% of poly(methacrylic acid) (molar mass: 100000 $g{\cdot}mol^{-1}$) (Polysciences, Inc.) in water distilled and purified with a milli-Q apparatus (Millipore). The molar mass of a repeat MAA unit is 87.1 $g{\cdot}mol^{-1}$, which corresponds to a molar concentration of 0.11 $mol{\cdot}L^{-1}$. The pH is adjusted by adding either a 0.1 M HCl solution (Sigma-Aldrich) or a 0.1 M NaOH solution (Sigma-Aldrich) and is measured with a pH-meter (pHM 250 ion analyser Meterlab, Radiometer Copenhagen) with a precision of 0.05 pH units. Oil-based PPO solutions are prepared by dissolution of 1 wt% of poly(propylene oxide) (molar mass: 4000 $g{\cdot}mol^{-1}$) (Sigma-Aldrich) in Mygliol 812 N (IMCD France/Sasol). Miglyol is a neutral oil consisting of caprylic/capric triglyceride (C8/C10 chains). The molar mass of a PO repeat unit is 58.1 $g{\cdot}mol^{-1}$, which corresponds to a molar concentration of 0.15 $mol{\cdot}L^{-1}$. We choose 1 wt% for both polymers to ensure an excess of polymer in bulk phases with respect to the interface, while being in the dilute regime (<3 wt%) to have a low viscosity solution, thus improving sensitivity to interfacial rheology measurements.

**The potentiometric measurements** were performed using a digital pH meter (pH M 250 ion analyzer, Meterlab, Radiometer Copenhagen) with a combined glass-calomel electrode. The electrode system was daily calibrated using three buffers (pH 4, 7 and 10). Potentiometric titrations were carried out with a sodium hydroxide (NaOH) solution (0.2 M), according to the PMAA concentration (0.1 $mol{\cdot}L^{-1}$). The measurements were made at 25°C, with constant stirring. pH values were recorded after sufficient stabilization. The dependence of the PMAA ionization degree α on pH was determined by using the potentiometric titration data, as:

$$\alpha = \frac{c_{NaOH} + c_{H+}}{c_{MAA}}, \quad (1)$$

where $c_{NaOH}$ and $c_{H+}$ are the molarities of added NaOH and free hydrogen ions, respectively, and $c_{MAA}$ is the molarity of monomer units (as each monomer unit bears one carboxylic group) [21]. The ionization degree defined above takes into account the fact that the molarity of hydroxide ions is very low for the cases of weak and moderate polymer ionization. $\alpha = 1$ corresponds to complete ionization.

The membrane thickness was measured in situ using an optical spectrometer V8E (Specim) assembled on an optical microscope (Olympus), with the focus strictly set at the interface where the membrane grows. The thickness was also measured ex situ using an optical profilometer (Microsurf 3D, Fogale nanotech) by transferring the membrane from the liquid to a glass slide. Emulsions were prepared in vials by gently pouring 6 mL of water-based solution and then 4 mL of oil-based solution.

**To probe the interfacial rheometry**, an AR-G2 rheometer (TA Instruments) is used with a Double-Wall-Ring geometry. The torque measurement resolution is 1 nN.m. The ring-shaped container is half-filled with approximately 21 mL of PMAA water solution until obtaining a flat interface pinned horizontally between the edges of the walls, so that meniscus deformation can be neglected. Then, the ring is carefully brought close to the interface and precisely positioned precisely to maintain a flat interface between the wall corner and the diamond-shaped corner of the ring. Finally, the rest of the container is slowly filled with the same volume of PPO solution in Miglyol. Measurements are controlled by TRIOS software (TA Instruments). A strain rate of 0.1% is imposed to ensure that all measurements are carried out in the linear regime. The presented measurements are all obtained one hour after the ring is deposited at the interface.

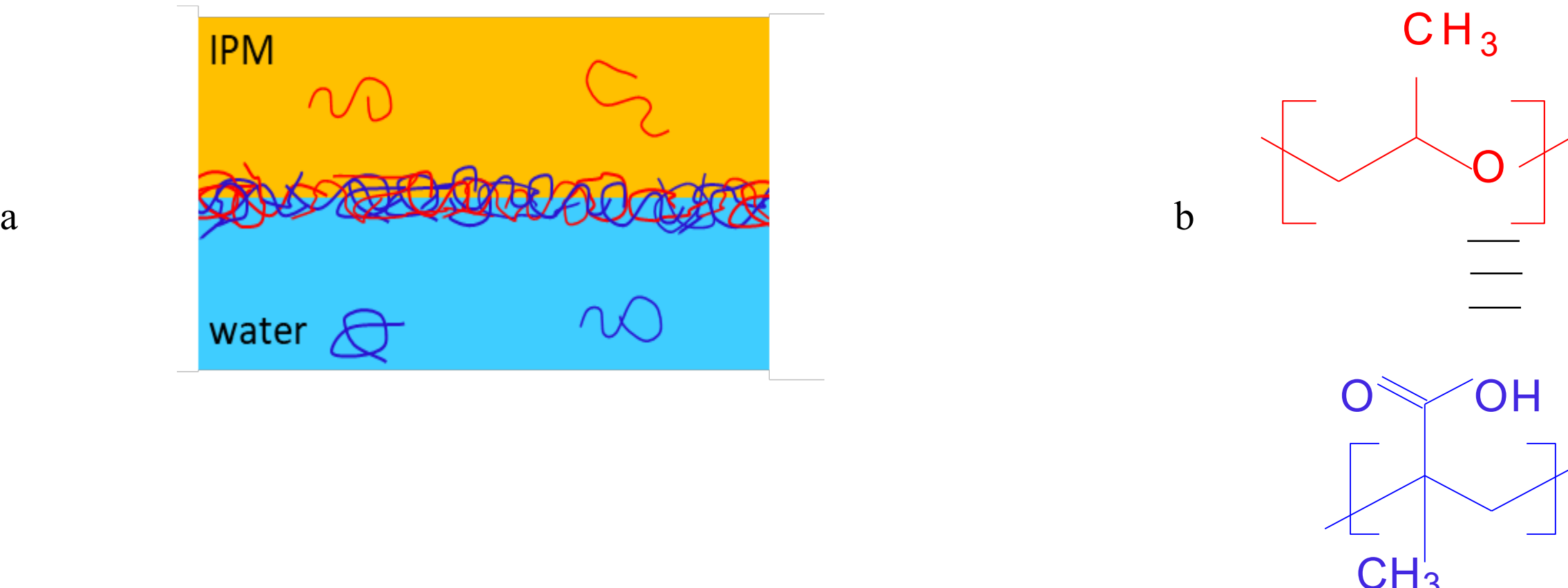

Figure 1. a. Schematic drawing of the interfacial assembly of PPO and PMAA polymer chains at the Miglyol water interface. b. PO repeat unit and MMA repeat unit interacting through hydrogen bonds

**Results and discussion**

As the pH increases from 3 to 5.1, the ionization degree of PMAA increases from 2 to 10% (inset in Figure 2). Hence, we expect that the ability of PMAA chains to establish hydrogen bonds with the PPO chains decreases with the pH, as recently shown [21]. Indeed, we observe that the thickness of the PMAA-PPO interfacial layer measured after one hour drops by almost a factor of 2, from 360 nm to 150 nm when the pH exceeds 4.5, which corresponds to an ionization degree $\alpha = 8\%$.

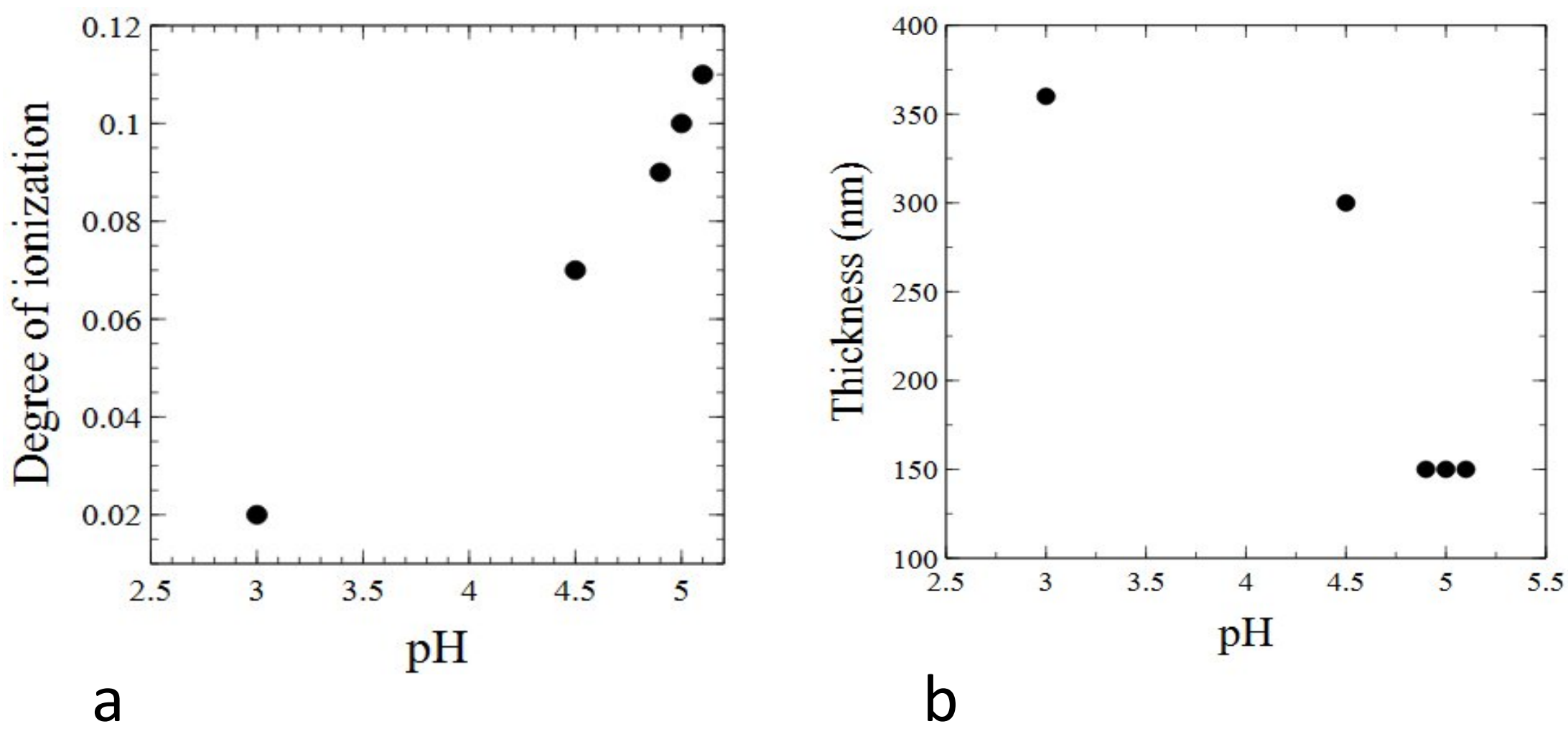

Figure 2 a. Degree of ionization $\alpha$ of PMAA as a function of pH. b. Thickness of the PPO-PMAA interfacial layer at the water-Miglyol interface after one hour as a function of pH. The PPO and PMAA are at 1 wt% in miglyol and water respectively.

Turning to the interfacial rheological properties, we report in Figure 3 the frequency sweeps of the storage and loss interfacial shear moduli G’ and G’’, measured at the interface between the PMAA and the PPO solution one hour after the establishment of contact between the two solutions. For a frequency of 0.01 rad.s$^{-1}$, $G'$ and $G''$ vary over 3 orders of magnitude for a pH ranging between 3 and 5 and are not even measurable for pH >= 5.25.

At pH = 3, corresponding to a degree of ionization of PMAA chains of $\alpha$ = 2 % the layer exhibits a viscoelastic behavior with a crossover angular frequency $\omega_c$ = 0.08 rad·s$^{-1}$. For $\omega < \omega_c$, $G' < G''$, meaning that the layer presents a fluid behavior. For $\omega > \omega_c$, $G' > G''$ and $G'$ becomes almost independent of frequency indicating that the layer presents an elastic behavior. As non-covalent hydrogen bonds between the PMAA and PPO chains control the rheology in the interfacial layer, it is expected that these interactions present a finite life time controlled by rates of association and dissociation [25-30]. When the deformation frequency exceeds the frequency associated with the time required for bound groups to dissociate, it is expected that these hydrogen bonds will behave as crosslinks between the polymer chains and provide an elastic resistance to the deformation [25-30]. In contrast, when the deformation frequency is lower than the frequency associated with the typical life time of these bonds, the layers are expected to present a viscous behavior consistently with what we observe [25-30].

When the pH is increased to pH = 4.5, corresponding to a degree of ionization of $\alpha$ = 8 %, the rheological curves $G'(\omega)$ and $G''(\omega)$ are shifted to higher frequencies. The cross-over frequency increases with the pH meaning that the typical life time of the hydrogen bonds is lower.

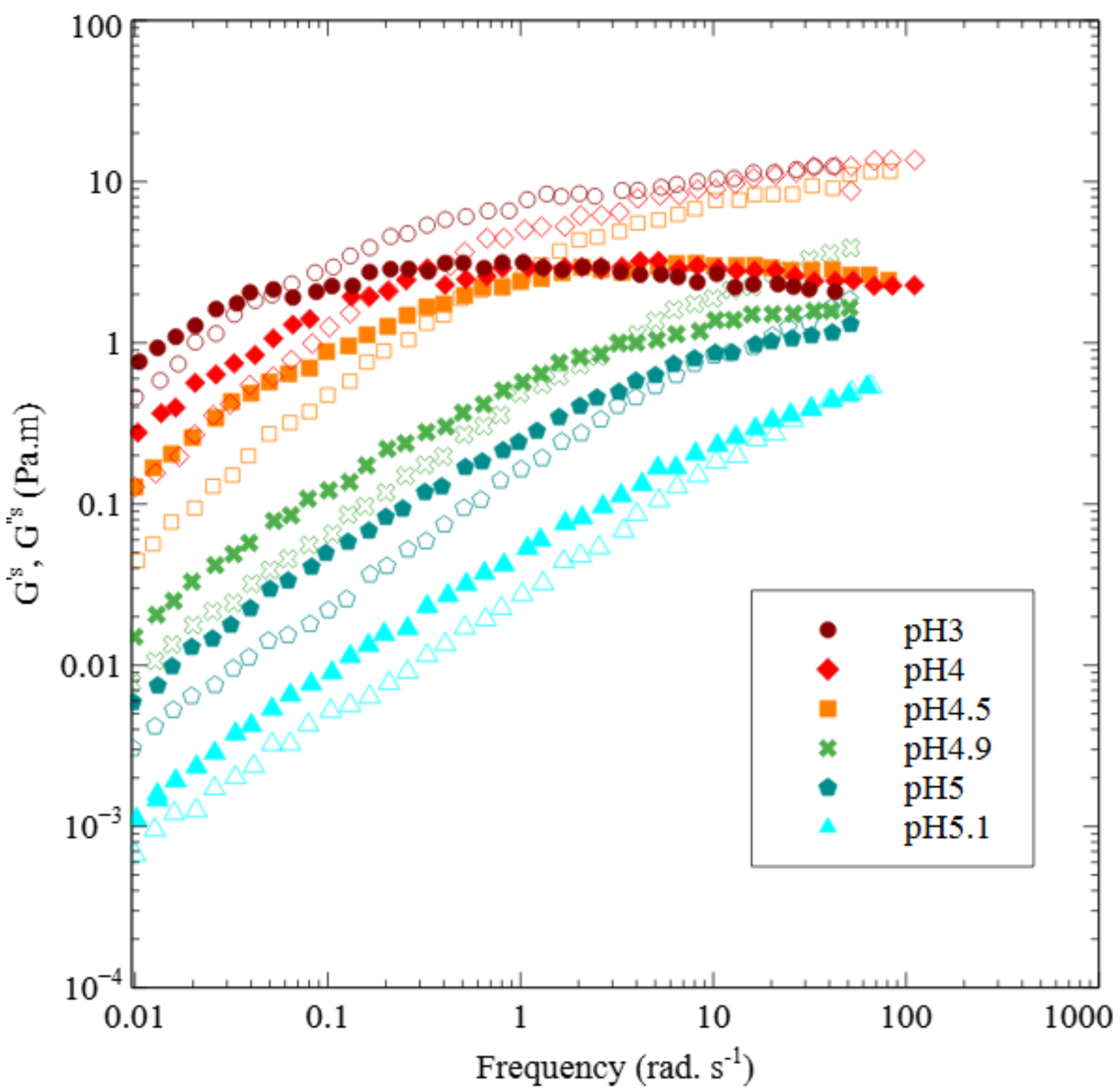

Figure 3. Frequency sweeps of the PPO-PMAA interfacial layers as a function of pH. The measurements are made for PPO solutions in mygliol and PMAA solutions in waterat 1 wt%. The measurements are performed one hour after the two solutions were put into contact. The deformation amplitude is 0.1 %. The open symbols represent $G''$ and filled symbols represent $G'$.

At pH = 5.1, $G'(\omega)$ and $G''(\omega)$ remain parallel over four decades of frequency and $G' \sim G'' \sim \omega^{0.7}$. Such behavior may be interpreted as a result of percolation of the gelled membrane, where there are just enough hydrogen bonds to form a weak viscoelastic network [30]. At a higher pH (pH > 5.25), no interfacial membrane was reported in our previous study where we used interferometry and atomic force microscopy to measure the membrane thicknesses [21], indicating that the polymers do not assemble at the interface. As a consequence, in this particular case, the rheological response is dominated by the bulk phases and is thus not reliable and thus not shown.

In order to gain further insight into the influence of pH on the interfacial rheology of the PMAA-PPO layers, we have reported in Figure 4a the values of the interfacial elastic modulus $G_c'^{s}$ at the crossover point as a function of pH. At pH = 3 and pH = 4.5, we measure a value of the crossover elastic modulus of $G_c'^s \sim 2$ Pa · m. $G_c'^s$ decreases to 0.5 Pa · m as the pH is increased above pH=4.9.

To eliminate the potential influence of the layer thickness on the measured values of $G_c'^s$, we renormalize the interfacial modulus values by the layer thickness, h, as reported in Figure 2, to obtain the values of the elastic modulus, i.e. $G_c' = G_c'^s/h$ . $G_c'$ has the unit of a stress (Pa) and should therefore represent the bulk shear elastic modulus of the membrane. For pH values up to 4.9 the values of $G_c'$ remain nearly constant around 6 MPa. When the pH increases to 5.1 a twofold decrease of the bulk shear modulus of the membrane is obtained. These results show that the bulk shear modulus presents a limited dependence with the pH, and that the decrease of the interfacial elastic modulus $G_c'^s$ observed for pH = 4.7 and 4.9 is mostly due to the reduction in the membrane thickness.

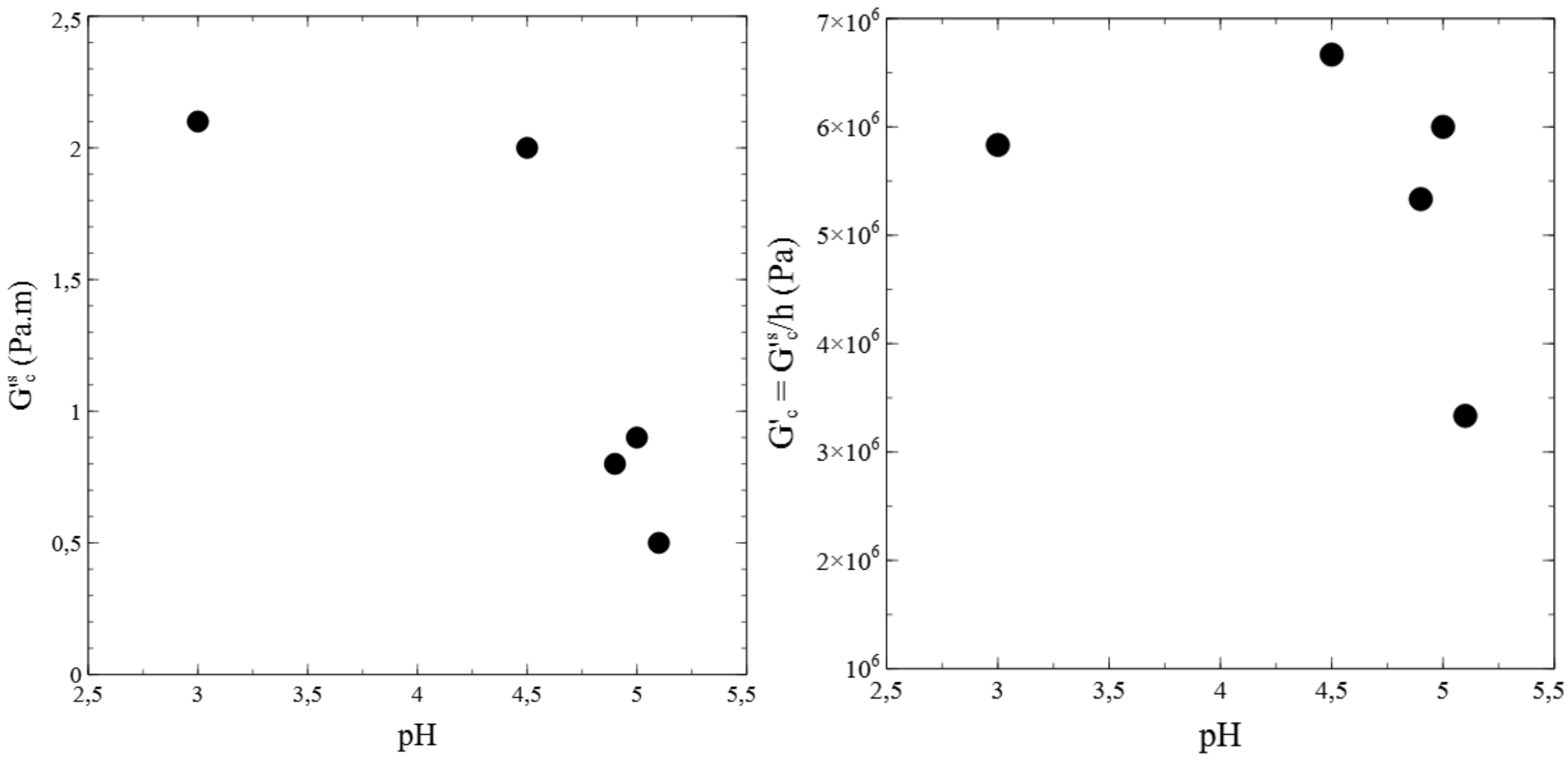

Figure 4. a. Values of the crossover interfacial modulus as a function of the ionization degree of PMAA. b. Values of $G_c' = G_c'^s/h$ obtained by rescaling the surface crossover modulus by the thickness of the interfacial layer.

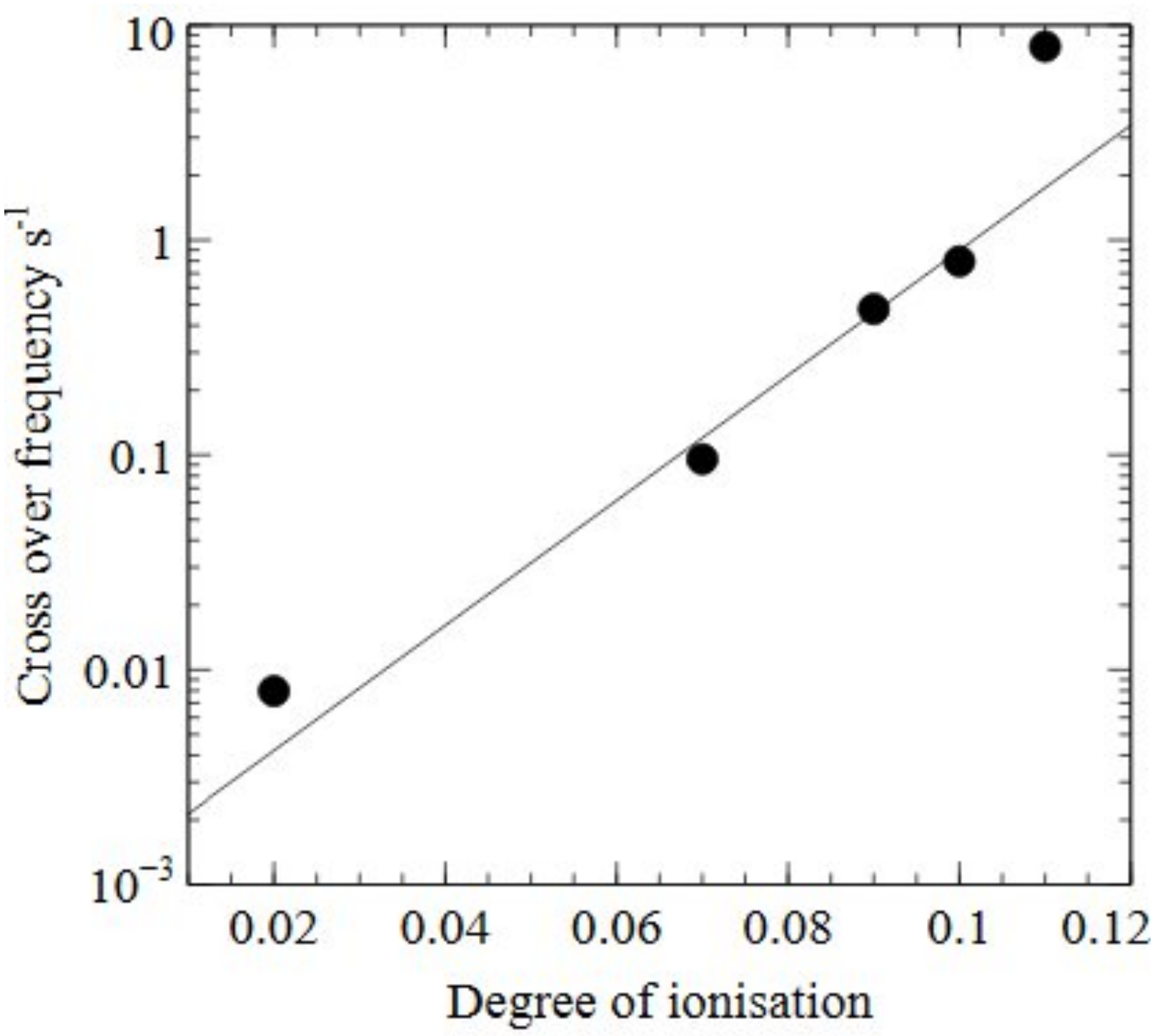

Figure 5. Crossover frequency of the rheology curves as a function of the degree of ionization of PMAA. The line is a fit of the data with an exponential function: $\omega_c = 0.001\ e^{69\ \alpha}$ rad·s$^{-1}$.

By contrast the pH has a much greater influence on the value of the crossover frequency, $\omega_c$. In Figure 5, it can be seen that $\omega_c$ increases by three orders of magnitude as the degree of ionization of PMAA rises from 0.01 to 0.11 which corresponds to a pH rise from 3 to 5.1. More precisely, the crossover frequency varies exponentially with the degree of ionization, i.e. $\omega_c \sim k \cdot e^{s \cdot \alpha}$ with k = 0.001 rad·s$^{-1}$ and s = 69, as shown in Figure 5.

In associative polymer networks, the cross over frequency obtained with rheological measurements is often related to the dissociation time of the reversible bonds [25-29] while the elastic modulus is related to the length or volume of the elastically active chains.

We can determine the volume of elastically active chains including both solvent and polymer as follows [31, 32]:

$$G'_c = \frac{k_B T}{V_s} \tag{4}$$

where $G'_c$ is the bulk shear elastic modulus at $\omega_c$.

For $G'_c = 6 \cdot 10^6$ Pa, we find $V_s \approx (0.9\ \mathrm{nm})^3$ which is consistent with a very short elastically-active subchain involving only a few monomer units. This result indicates that there is a high density of hydrogen bond interactions between PPO and PMAA chains in the membrane which depends weakly on the pH.

To account for the exponential variation of the cross over frequency with the pH we hypothesize that the PPO chains behave as multifunctional reversible crosslinkers that bind to several PMAA chains through hydrogen bonds as schematized in Figure 6. With such picture, the cross-over frequency we measure is related to the probability of one PPO chain to dissociate all of its links with one of these PMAA chains. Once the PPO and PMMA chains have detached, their free parts can relax and switch to other positions.

Assuming that the experimentally-measured crossover frequency is related to the time required for detaching all the links between two bounded PPO and PMAA chains, we provide an estimate of the number $N_c$ of such links and the fraction of monomers involved in hydrogen interactions. Let us assume an Arrhenius-type dependence for the individual dissociation dynamics. Then, the rate of dissociation of two bounded chains reads:

$$w_c = w_0 e^{-E/k_B T} = w_0 e^{-N_c \varepsilon/k_B T} = w_0 e^{-s(1-\alpha)}, \quad (5)$$

where E is the interaction energy of two bounded chains, $k_B$ is the Boltzmann constant, T is the temperature, $w_0$ is a prefactor, s is the exponent in the exponential law defined earlier, and ε is the interaction energy of a single hydrogen bond between one MAA unit and one PO unit.

Denoting $\varepsilon = n k_B T$, where n = 5 to 10 from Ref. [33], we deduce from Eq. (2) that the number of links between two bounded PPO and PMAA chains decreases linearly with increasing ionization degree α:

$$N_c = \frac{s}{n}(1-\alpha) \quad . \quad (6)$$

From Figure 5, we obtain s = 69. For α = 1, we have thus $N_c \approx s/n = 7$ to 14. Therefore, there are 7 to 14 non-covalent bonds between one given PPO and one given PMAA chain, and each PPO chain binds to several PMAA chains (Fig. 6). As the pH increases, the number of charged metacrylate groups on the PMAA chains increases. These metacrylate groups cannot bind to PO units through hydrogen bonds. As a result, the number, $N_c$, of PO units on one PPO chain binding to a given PMAA chain decreases. As the dissociation rate, w required for one PPO chain to detach all its hydrogen bonds established with PMAA chains increases exponentially with $N_c$, this results in a strong increase of the cross-over frequency measured from interfacial rheology measurements.

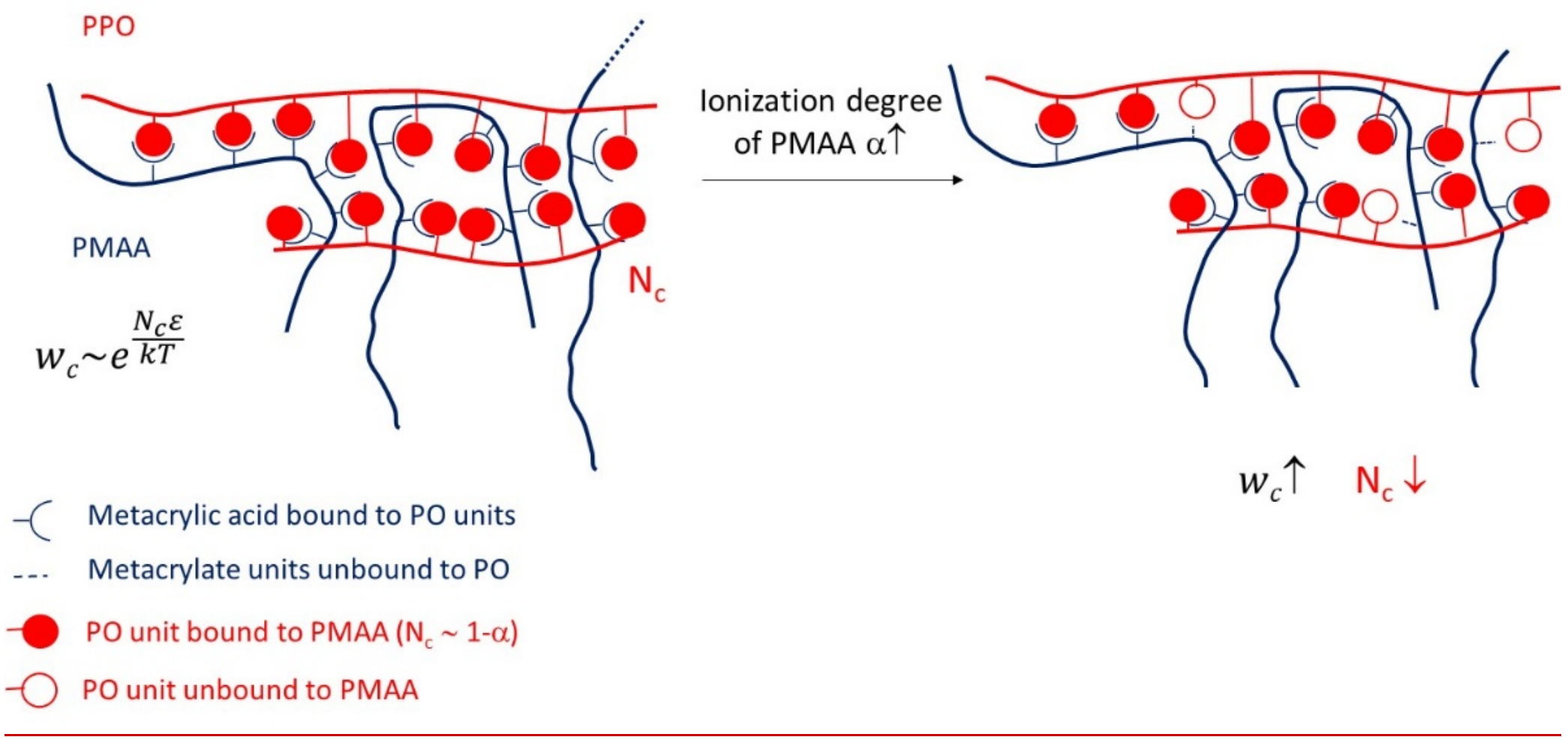

Figure 6. Schematic representation of the network between PMAA and PPO chains. For each PPO chain, several units are involved in hydrogen bonding which enables it to bridge several PMAA chains

## Conclusion

We have measured the interfacial shear rheological properties of interfacial layers assembled through reversible hydrogen interactions between PPO and PMAA. These layers exhibit a viscoelastic behavior which depends strongly on pH. Varying the pH enables us to tune the values of the elastic modulus only by a factor two while the crossover frequency increases exponentially with the degree of ionization of the PMAA chains. To describe our results, we propose a simple picture in which one PPO chain behaves as a multipoint crosslinker bridging several PMAA chains. With this picture, we deduce from our experimental results reasonable values for the number of PPO monomers involved in such interfacial layers.

## Acknowledgements

The authors acknowledge financial support from the Agence Nationale de la Recherche under Interpol (ANR-12-JS08-0007), EMetBrown (ANR-21-ERCC-0010-01), Softer (ANR21-CE06-0029), and Fricolas (ANR-21-CE06-0039) grants. The authors thank Julie Brun for performing some of the experiments. They also acknowledge financial support from the European Union through the European Research Council under EMetBrown (ERC-CoG-101039103) grant. Views and opinions expressed are however those of the authors only and do not necessarily reflect those of the European Union or the European Research Council. Neither the European Union nor the granting authority can be held responsible for them. Finally, the authors thank the Soft Matter Collaborative Research Unit, Frontier Research Center for Advanced Material and Life Science, Faculty of Advanced Life Science at Hokkaido University, Sapporo, Japan, and the CNRS International Research Network between France and India on "Hydrodynamics at small scales: from soft matter to bioengineering".